\documentclass{PoS}
\usepackage{amsmath,bm}
\usepackage{xcolor}
\usepackage[utf8]{inputenc}
\usepackage{amsfonts}
\usepackage{amssymb}
\usepackage[numbers,sort&compress]{natbib}
\usepackage{tikz}
\usetikzlibrary{patterns}
\usetikzlibrary{decorations}
\usepackage{comment}

\newcommand{\ba}{\begin{array}{c}}
\newcommand{\bat}{\begin{array}{cc}}
\newcommand{\ea}{\end{array}}

\title{A complete update of $\varepsilon'/\varepsilon$ in the Standard Model}

\ShortTitle{A complete update of $\varepsilon'/\varepsilon$ in the Standard Model}

\author{V. Cirigliano\\
        Theoretical Division, Los Alamos National Laboratory, Los Alamos, NM 87545, USA\\
        E-mail: \email{cirigliano@lanl.gov}}
        
\author{\speaker{H. Gisbert}\\
        Fakult\"at Physik, TU Dortmund, \\
        Otto-Hahn-Str.4, D-44221 Dortmund, Germany\\
        E-mail: \email{hector.gisbert@tu-dortmund.de}}
        
\author{A. Pich\\
        Departament de F\'{\i}sica Te\`orica, IFIC, CSIC --- Universitat de Val\`encia \\ 
        Edifici d'Instituts de Paterna, Apt. Correus 22085, E-46071 Val\`encia, Spain\\
        E-mail: \email{pich@ific.uv.es}}
        
\author{A. Rodr\'iguez-S\'anchez\\
       Department of Astronomy and Theoretical Physics, Lund University, S\"{o}lvegatan 14A, SE 223-62 Lund, Sweden\\
       E-mail: \email{antonio.rodriguez@thep.lu.se}}

\abstract{The recent release of improved lattice data has revived again the interest on precise theoretical calculations of the direct CP-violation ratio $\varepsilon'/\varepsilon$. We present a complete update of the Standard Model prediction [1,2], including a new re-analysis of isospin-breaking corrections which are of vital importance in the theoretical determination of this observable. The Standard Model prediction, $\mathrm{Re} (\epsilon'/\epsilon) = (14\pm 5)\cdot 10^{-4}$, turns out to be in good agreement with the experimental measurement.}

\FullConference{
European Physical Society Conference on High Energy Physics - EPS-HEP2019 -\\
			10-17 July, 2019\\
			Ghent, Belgium}

\begin{document}
\vspace{-0.25cm}
\section{Introduction}
\vspace{-0.25cm}
The matter-antimatter asymmetry in our Universe requires the violation of the CP symmetry. Although it has been observed in $B$, $D$ and $K$ systems, the amount of CP violation in the Standard Model (SM) is too low to reproduce the observed asymmetry, hence new sources of CP violation are needed to explain this large imbalance. The CP-violating ratio $\varepsilon'/\varepsilon$ represents a fundamental test for our understanding of this phenomena. In the SM, this observable is proportional to those Cabibbo-Kobayashi-Maskawa (CKM) matrix elements that account for the violation of this symmetry, therefore any new source of CP violation should have a direct impact on this ratio.

The different sources of CP violation in $K$ decays are parametrized by $\varepsilon'$ and $\varepsilon$, which are related with the branching ratios of the $K_L$ and $K_S$ decays into two pions,
\begin{align}
        \eta_{+-}\equiv\frac{A(K_L\to\pi^+\pi^-)}{A(K_S\to\pi^+\pi^-)}=\varepsilon+\varepsilon'~,\qquad  \eta_{00}\equiv\frac{A(K_L\to\pi^0\pi^0)}{A(K_S\to\pi^0\pi^0)}=\varepsilon-2\,\varepsilon'~.
\end{align}
The dominant effect from CP violation in $K$ mixing is contained in $\varepsilon$, and its experimental value is a per-mill effect $|\varepsilon|\,=\,(2.228\pm 0.011)\cdot 10^{-3}$ \cite{Tanabashi:2018oca}. In the case of $\varepsilon'$, which depends on the difference between $\eta_{+-}$ and $\eta_{00}$, the effect is tinier. Its experimental average \cite{Batley:2002gn,Abouzaid:2010ny} 
\begin{align}
\text{Re}(\varepsilon'/\varepsilon)_{\text{exp}}\,=\,\left(16.6\pm 2.3\right)\cdot 10^{-4}~,\label{eq:expeps}    
\end{align}
clearly demonstrates the existence of direct CP violation in $K$ decays. In addition, its small size makes it particularly sensitive to new sources of CP violation, providing a formidable way to search for physics beyond the SM.

\vspace{-0.25cm}
\section{The fingerprints of $K\to\pi\pi$ decays}
\vspace{-0.25cm}

\begin{figure}
    \centering
    \includegraphics[scale=0.25]{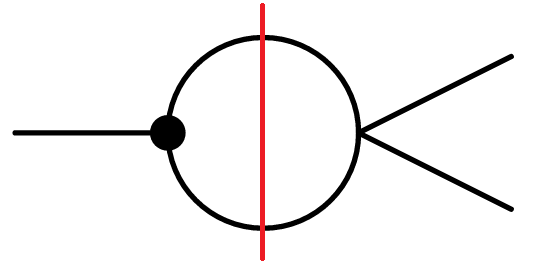}
    \caption{One-loop contribution to $K\to\pi\pi$ with its absorptive cut in red.}
    \label{fig:feynmandiagram}
\end{figure}

In this section, we explore the dynamical features of $K\to\pi\pi$ decays, taking into account the experimental data. This goal requires to adopt the usual isospin decomposition of the physical amplitudes \cite{Cirigliano:2003gt}
\begin{align}
    A[K^0\to\pi^+\pi^-]\, &=\, A_0\,\text{e}^{i\,\chi_0}\,+\,\frac{1}{\sqrt{2}}\,A_2\,\text{e}^{i\,\chi_2}\, =\,\mathcal{A}_{1/2}\,+\,\frac{1}{\sqrt{2}}\, (\mathcal{A}_{3/2}+\mathcal{A}_{5/2})~,\nonumber\\
    A[K^0\to\pi^0\pi^0]\, &=\, A_0\,\text{e}^{i\,\chi_0}\,-\,\sqrt{2}\,A_2\,\text{e}^{i\,\chi_2}\, =\,\mathcal{A}_{1/2}\,-\,\sqrt{2}\,(\mathcal{A}_{3/2}+\mathcal{A}_{5/2})~,\label{eq:decomp}\\
    A[K^+\to\pi^+\pi^0]\, &=\,\frac{3}{2}\,A_2^+\,\text{e}^{i\,\chi_2^+}\, =\, \frac{3}{2}\,\left(\mathcal{A}_{3/2}\,-\,\frac{2}{3}\,\mathcal{A}_{5/2}\right)~,\nonumber
\end{align}
where $\mathcal{A}_{1/2}\equiv A_0 \text{e}^{i \chi_0}$, $\mathcal{A}_{3/2}+\mathcal{A}_{5/2}\equiv A_2 \text{e}^{i \chi_2}$ and $\mathcal{A}_{3/2}-(2/3)\mathcal{A}_{5/2}\equiv A_2^+ \text{e}^{i \chi_2^+}$. 
In the isospin limit, $A_0$ and $A_2=A_2^+$ are the decay amplitudes into $(\pi\pi)_{0,2}$ states, and $\chi_I$ can be identified with the S-wave $\pi\pi$ scattering phase shifts $\delta_{I}$. In the CP-conserving limit, the amplitudes $A_I$ are real and positive. Using Eqs.~\eqref{eq:decomp} and the measured $K\to\pi\pi$ branching ratio, one obtains \cite{Antonelli:2010yf} 
\begin{align}
    A_0=(2.704 \pm 0.001)\cdot 10^{-7}\,\text{GeV}&,
    \qquad \quad
    A_2=(1.210 \pm 0.002)\cdot 10^{-8}\,\text{GeV},\label{eq:phenonumbers1}\\
    \chi_0-\chi_2&=(47.5 \pm 0.9)^\circ.\label{eq:phenonumbers2}
\end{align}
When CP violation is turned on, the amplitudes $A_{0,2}$ and $A_2^+$ acquire imaginary parts, and $\varepsilon'$ can be written to first order in CP violation as
\begin{align}
\varepsilon'=-\frac{i}{\sqrt{2}}\,\text{e}^{i\,(\chi_2-\chi_0)}\,\omega\,\left[\frac{\text{Im}A_0}{\text{Re}A_0}-\frac{\text{Im}A_2}{\text{Re}A_2}\right]\,=\,-\frac{i}{\sqrt{2}}\,\text{e}^{i\,(\chi_2-\chi_0)}\,\omega\,\frac{\text{Im}A_0}{\text{Re}A_0}\,\left(1-\frac{1}{\omega}\frac{\text{Im}A_2}{\text{Im}A_0}\right)~. \label{eq:epsp}   
\end{align}
Taking into account Eqs.~\eqref{eq:phenonumbers1} and \eqref{eq:phenonumbers2} together with Eq.~\eqref{eq:epsp}, we can easily study the impact of the $K\to\pi\pi$ dynamical properties  on $\varepsilon'$:
\begin{itemize}
    \item Eqs.~\eqref{eq:phenonumbers1} exhibit the well-known ``$\Delta I =1/2$ rule'', i.e., a large enhancement of the isoscalar isospin amplitude with respect the isotensor one, 
    \begin{align}
        \omega^{-1}\equiv\frac{\text{Re}A_0}{\text{Re}A_2}\approx 22~,\label{eq:isorule}
    \end{align}
    which directly implies a strong suppression of $\varepsilon'$. In addition, any small isospin-breaking correction to the ratio $\text{Im}A_2/\text{Im}A_0$ is enhanced by the factor $\omega^{-1}$ in Eq.~\eqref{eq:epsp}.
    \item Furthermore, Eq.~\eqref{eq:phenonumbers2} shows that the S-wave $\pi\pi$ re-scattering generates a large phase-shift difference between the $I = 0$ and $I = 2$ partial waves, which implies \cite{Gisbert:2018tuf}
    \begin{align}
        \text{Abs}(\mathcal{A}_{1/2}/\mathcal{A}_{3/2})\approx\text{Dis}(\mathcal{A}_{1/2}/\mathcal{A}_{3/2})~.
    \end{align}
    Thus, the absorptive contribution to this ratio is of the same size as the dispersive one. A good theoretical control of both contributions is then mandatory to obtain a reliable prediction for $\text{Re}(\varepsilon'/\varepsilon)$. 
    
    \item The presence of absorptive contributions is a direct consequence of unitarity, which becomes specially relevant for the isoscalar amplitude
    $\mathcal{A}_{1/2}\equiv A_0\,\text{e}^{i\,\delta_0}\,=\,\text{Dis}(\mathcal{A}_{1/2})\,+\,i\,\text{Abs}(\mathcal{A}_{1/2}) $.
    Using the known value of the $I=0$ phase shift, $\delta_0 = (39.2\pm 1.5)^\circ$ \cite{Colangelo:2000jc}, one immediately obtains
    \begin{align}
        A_0\,\equiv\, |\mathcal{A}_{1/2}|\,=\,\text{Dis}(\mathcal{A}_{1/2})\,\sqrt{1+\tan^2\delta_0}\approx 1.3\times\text{Dis}(\mathcal{A}_{1/2})~.\label{eq:A0uni}
    \end{align}
    Therefore, the absorptive contribution increases the numerical size of $A_0$ by 30\%. 
    
    \item The absorptive amplitudes are generated by intermediate on-shell pions, through the
    Feynman-diagram topology depicted in Figure~\ref{fig:feynmandiagram}. The dispersive and absorptive loop contributions are related by analyticity. A large absorptive contribution implies a large dispersive loop correction. 
    
\end{itemize}

\vspace{-0.25cm}
\section{Current  estimate of $\varepsilon'/\varepsilon$ from lattice QCD}
\vspace{-0.25cm}

In 2015, the RBC-UKQCD collaboration published their first estimate of $\varepsilon'/\varepsilon$
\cite{Blum:2015ywa,Bai:2015nea}:
\begin{align}
\text{Re}(\varepsilon'/\varepsilon)_{\text{lattice}}\,=\,\left(1.4\pm 6.9\right)\cdot 10^{-4}~.\label{eq:epsplattice}    
\end{align}
This result is consistent with zero and shows a clear discrepancy of 2.1$\,\sigma$ with the experimental value given in Eq.~\eqref{eq:expeps}. The disagreement has triggered many analyses of possible new physics contributions in order to explain the apparent anomaly. However, one should realize the technical limitations of this lattice estimate. For example, the phase shifts $\delta_{0,2}$ play a crucial role in the lattice determination of $\varepsilon'/\varepsilon$ and provide a quantitative test of the obtained result. While the extracted $I=2$ phase shift is only 1$\,\sigma$ away from its experimental value, the RBC-UKQCD collaboration finds a $I = 0$ phase shift that disagrees with its experimental value by 2.9$\,\sigma$. This discrepancy is much larger than the one exhibited by their $\varepsilon'/\varepsilon$ result.

Therefore, it is still premature to derive strong implications from the 2015 RBC-UKQCD lattice data, since the important effects of $\pi\pi$ re-scattering are still not well reproduced in
the $I=0$ amplitude. Efforts towards a better lattice determination are under way \cite{Fu:2013ffa,Kelly:2019yxg}.  

Notice that the recent claims of an $\varepsilon'/\varepsilon$ anomaly, from groups using analytical methods \cite{Buras:2015yba}, are based on simplified calculations which either use the RBC-UKQCD matrix elements (with somewhat smaller uncertainties) or adopt model-dependent $K\to\pi\pi$ amplitudes without any absorptive components, missing completely the important $\pi\pi$ re-scattering corrections.

\vspace{-0.25cm}
\section{Multi-scale framework}\label{sec:framework}
\vspace{-0.25cm}

Due to the presence of widely separated mass scales ($M_\pi<M_K\ll M_W$), the theoretical description of the $K\to\pi\pi$ decays requires the use of two different effective field theories (EFTs). 
Above the electroweak scale $M_W$, all flavour-changing processes are described in terms of quarks, leptons and gauge bosons. We can apply the renormalization group equations and the operator product expansion to go down to low-energy scales ($\sim\,$1 GeV), integrating out all the heavy particles in the way. Finally, one obtains the effective $\Delta\,S=1$ short-distance Lagrangian \cite{Buchalla:1995vs}
\begin{align}
    \mathcal{L}_{\text{eff}}^{\Delta\, S=1}\,=\,-\frac{G_F}{\sqrt{2}}\,V_{ud}\,V_{us}^*\,\sum_{i=1}^{10}\,C_i(\mu)\,Q_i(\mu)~,\label{eq:shortL}
\end{align}
which is a sum of local four-fermion operators that are weighted by the Wilson coefficients, $C_i(\mu)\,=\,z_i(\mu)+\tau\,y_i(\mu)$. The dependence on the CKM matrix elements is carried by the global $V_{ud}\,V_{us}^*$ factor and the parameter $\tau\equiv-V_{td}V_{ts}^*/(V_{ud}V_{us}^*)$ that contains the CP-violating phase.
The information on the heavy masses has been absorbed into the Wilson coefficients $C_i(\mu)$, which are known at next-to-leading-order (NLO) \cite{Buras:1991jm,Buras:1992tc,Buras:1992zv,Ciuchini:1993vr}. Some next-to-next-to-leading-order (NNLO) corrections \cite{Buras:1999st,Gorbahn:2004my} are already known and efforts towards a complete calculation at the NNLO are currently under way \cite{Cerda-Sevilla:2016yzo}.

Below the resonance region where the physics of study is defined in terms of Goldstone bosons ($\pi$, $K$, $\eta$), one can use symmetry considerations in order to build an EFT valid in this non-perturbative regime. Chiral Perturbation Theory ($\chi$PT) provides a formidable theoretical framework to describe the pseudoscalar-octet dynamics as a perturbative expansion in powers of momenta and quark masses over the chiral symmetry-breaking scale $\Lambda_\chi$~. Using the chiral symmetry, we can build all the allowed operators with exactly the same symmetry properties as the short-distance Lagrangian~\eqref{eq:shortL}. To lowest order, the chiral realization of $\mathcal{L}_{\text{eff}}^{\Delta\, S=1}$ contains three operators
\begin{align}
\mathcal{L}_2^{\Delta S =1}\,=\,G_8\,\mathcal{L}_8\,+\,G_{27}\,\mathcal{L}_{27}\,+\,G_8\,g_{\text{ewk}}\,\mathcal{L}_{\text{ewk}}~,\label{eq:longL}  
\end{align}
with their respective low-energy couplings (LECs) $G_8$, $G_{27}$ and $G_8\,g_{\text{ewk}}$ encoding all quantum information from high-energy scales \cite{Cirigliano:2011ny}. 
The determination of these LECs requires to perform a matching between both Lagrangians~\eqref{eq:shortL} and \eqref{eq:longL} in a common region of validity. However, performing consistently this matching is a very challenging task that still remains unsolved. The large-$N_C$ limit provides a partial solution to this problem. In this limit, the T-product of two colour-singlet currents factorizes and, since we have a well-known representation for these currents in $\chi$PT, the matching can be done at leading order in the $1/N_C$ expansion. It is important to remind that the missing NLO contributions to the matching are not enhanced by any large logarithms. 

\vspace{-0.25cm}
\section{Isospin-breaking corrections to $\varepsilon'/\varepsilon$}
\vspace{-0.25cm}

Eqs.~\eqref{eq:epsp} and \eqref{eq:isorule} exhibit the important role of isospin-breaking effects in $\varepsilon'/\varepsilon$. Including these corrections, $\text{Re}(\varepsilon'/\varepsilon)$ can be written as \cite{Cirigliano:2003nn,Cirigliano:2003gt}
\begin{align}
\mathrm{Re}\Bigl(\frac{\varepsilon'}{\varepsilon}\Bigr) \, = \, - \frac{\omega_+}{\sqrt{2}\, |\varepsilon|}  \, \left[
\frac{\mathrm{Im}\: A_{0}^{(0)} }{ \mathrm{Re}\: A_{0}^{(0)} }\,
\left( 1 - \Omega_{\rm eff} \right) - \frac{\mathrm{Im}\: A_{2}^{\rm emp}}{ \mathrm{Re}
  \: A_{2}^{(0)} } \right]  , \;\;
\end{align}
where the superscript $(0)$ denotes the isospin limit, $\mathrm{Im}\: A_{2}^{\rm emp}$ contains the $I=2$ contribution from the electromagnetic penguin operators and $\omega_+ \equiv \mathrm{Re}\: A_{2}^{+}/\mathrm{Re} \: A_{0}$. The parameter $\Omega_{\rm eff}$ contains the isospin-breaking corrections. 
Implementing the current improvements on the inputs that enter in this parameter, we have updated the $\Omega_{\rm eff}$ prediction with the result \cite{Cirigliano:2019cpi} 
\begin{align}\label{eq:Omega_eff_res}
\Omega_{\text{eff}}\:=\:(11.0\,{}^{\,+\,9.0}_{\,-\,8.8})\cdot 10^{-2}\, ,
\end{align}
which agrees within errors with the previous determination \cite{Cirigliano:2003nn,Cirigliano:2003gt} but has a larger central value.

\vspace{-0.25cm}
\section{Strong cancellation in simplified analyses}
\vspace{-0.25cm}

The CP-odd amplitudes $\text{Im}\,A_{0,2}$ are mainly dominated by $(V-A)\times(V+A)$ operators because they have a chiral enhancement that can be easily estimated in the large-$N_C$ limit. Due to the size of $y_i(\mu)$, it is a good numerical approximation to consider only $Q_6$ and $Q_8$ and neglect the contributions to $\text{Im}\,A_{0,2}$ from other operators. With this rough estimation, one obtains \cite{Gisbert:2017vvj} 
\begin{align}
    \mathrm{Re}(\varepsilon'/\varepsilon) \, \approx \, 
2.2\cdot 10^{-3}
\left\{ B_6^{(1/2)} \left( 1 - \Omega_{\rm eff} \right) - 0.48\, B_8^{(3/2)}
\right\}  ,\;\label{eq:naiveeps}
\end{align}
where $B_6^{(1/2)}$ and $B_8^{(3/2)}$ parametrize the deviations of the true hadronic matrix elements from their large-$N_C$ approximations $B_6^{(1/2)}=B_8^{(3/2)}=1$, which do not include any absorptive contribution. Taking $\Omega_{\rm eff}=0.11$ \cite{Cirigliano:2019cpi}, Eq.~\eqref{eq:naiveeps} gives $\mathrm{Re}(\varepsilon'/\varepsilon)\approx 0.9\cdot 10^{-3}$ at $N_C\to\infty$; the same order of magnitude as its experimental value in Eq.~\eqref{eq:expeps}. In contrast, with the values adopted in Ref.~\cite{Buras:2015yba}, $B_6^{(1/2)}=0.57$, $B_8^{(3/2)}=0.76$ and $\Omega_{\rm eff}=0.15$, one gets $\mathrm{Re}(\varepsilon'/\varepsilon)\approx 2.6\cdot 10^{-4}$, one order of magnitude smaller than \eqref{eq:expeps}. Clearly, with this choice of $B_{6,8}$ parameters, the simplified approximation in Eq.~\eqref{eq:naiveeps} suffers a strong cancellation between the different contributions.

\renewcommand*{\thefootnote}{\fnsymbol{footnote}}
\setcounter{footnote}{0}

We can go one step further and include naively the chiral loop corrections \cite{Pallante:1999qf,Pallante:2000hk,Pallante:2001he,Cirigliano:2003gt} (Figure~\ref{fig:feynmandiagram}). These contributions are mainly dominated by $\Delta_L\mathcal{A}_{1/2}^{(8)}$ and $\Delta_L\mathcal{A}_{3/2}^{(g)}$ , which imply the following shifts, $B_6^{(1/2)}\to |1+\Delta_L\mathcal{A}_{1/2}^{(8)}|\,B_6^{(1/2)}\approx 1.35\,B_6^{(1/2)}$ and $B_8^{(3/2)}\to |1+\Delta_L\mathcal{A}_{3/2}^{(g)}|\,B_8^{(3/2)}\approx 0.54\,B_8^{(3/2)}$, in Eq.~\eqref{eq:naiveeps}. With this shifts in mind, we can again estimate $\mathrm{Re}(\varepsilon'/\varepsilon)$ for different setups, see Table~\ref{tab:setupeps}.
We can observe that the chiral loop corrections destroy the strong numerical cancellation in Eq.~\eqref{eq:naiveeps}, yielding results of the same order of magnitude as the experimental measurement.

{\renewcommand{\arraystretch}{1.2}%
\begin{table}[htb!]
\centering
\begin{tabular}{|l|c|c|l|}
 \hline
Set-up & $B_6^{(1/2)}$ & $B_8^{(3/2)}$  &  $\mathrm{Re}(\varepsilon'/\varepsilon)$ \\\hline
Large $N_C\,+\,$FSI$_{I=0,2}$& 1.35& 0.54 &   $2.1\cdot 10^{-3}$ \\\hline
Large ${N_C|}_{I=0}$ + LQCD$_{I=2}$ $+\,$FSI$_{I=0}$ & 1.35& 0.76& $1.7\cdot 10^{-3}$\\ \hline
LQCD $+\,$FSI$_{I=0}$ \footnotemark  &0.77 & 0.76& $0.6\cdot 10^{-3}$ \footnotemark[2]\\ \hline
\end{tabular}
\caption{Naive estimates of $\text{Re}(\varepsilon'/\varepsilon)$, including some final-state interactions (FSI) in Eq.~\eqref{eq:naiveeps}.}
\label{tab:setupeps}
\end{table}}
\footnotetext{This setup suffers from a double counting of the $\chi$PT and lattice FSI with $I=0$.}
\footnotetext[2]{The resulting numerical estimate corresponds to the central value in Ref.~\cite{Aebischer:2019mtr}.}

\vspace{-0.25cm}
\section{Standard Model prediction for $\varepsilon'/\varepsilon$ in $\chi$PT}
\vspace{-0.25cm}

With the theoretical framework presented in Section~\ref{sec:framework}, which includes all four-fermion operators (not only $Q_6$ and $Q_8$), the full 1-loop $\chi$PT contributions and the new updated isospin-breaking corrections given by Eq.~\eqref{eq:Omega_eff_res}, our SM prediction for $\text{Re}\left(\epsilon'/\epsilon\right)$~\cite{Cirigliano:2019cpi},
\begin{align}\label{eq:epsp_prediction}
\text{Re}\left(\epsilon'/\epsilon\right)\,&=\,\left(13.8\pm 1.3_{\,\gamma_5}\pm 2.5_{\,\text{LECs}}\pm 3.5_{\,1/N_C}\right)\cdot 10^{-4}\:=\:\left(14\,\pm\,5\right)\cdot 10^{-4} \, ,
\end{align}
is in excellent agreement with the experimental world average in Eq.~\eqref{eq:expeps}. Eq.~\eqref{eq:epsp_prediction} displays the three different sources of uncertainty in $\text{Re}\left(\epsilon'/\epsilon\right)$. The first error reflects the choice of scheme for $\gamma_5$. The second error originates from the input values of the strong LECs $L_{5,7,8}$. The last error parametrizes our ignorance about $1/N_{C}$-suppressed contributions in the matching region which have been estimated very conservatively through the variation of $\mu_{\rm SD}$ and $\nu_{\chi}$ in the intervals $ [0.9,1.2] \, \mathrm{GeV}$ and $[0.6,1]$~GeV, respectively. Further details can be found in Ref.~\cite{Cirigliano:2019cpi}.

\vspace{-0.25cm}
\section*{Acknowledgements}
\vspace{-0.25cm}

We want to thank the organizers for their effort to make this conference such a successful event. This work has been supported in part by the Spanish Government and ERDF funds from the EU Commission [grant FPA2017-84445-P], the Generalitat Valenciana [grant Prometeo/2017/053], the Spanish Centro de Excelencia Severo Ochoa Programme [grant SEV-2014-0398], the Swedish Research Council [grants 2015-04089  and  2016-05996]  and  by  the  European  Research Council (ERC) under the EU Horizon 2020 research and innovation programme (grant 668679). The work of H.G. is supported by a FPI doctoral contract [BES-2015-073138], funded by the Spanish Ministry of Economy, Industry and Competitiveness and the Bundesministerium f\"ur Bildung und Forschung (BMBF). V.C. acknowledges support by the US DOE Office of Nuclear Physics.


\begin{thebibliography}{99}
  
\bibitem{Cirigliano:2019cpi}
  V.~Cirigliano, H.~Gisbert, A.~Pich and A.~Rodr\'iguez-S\'anchez,
  arXiv:1911.01359 [hep-ph].

\bibitem{Gisbert:2017vvj}
  H.~Gisbert and A.~Pich,
  Rept.\ Prog.\ Phys.\  {\bf 81} (2018) no.7,  076201.


\bibitem{Tanabashi:2018oca}
  M.~Tanabashi {\it et al.} [Particle Data Group],
  Phys.\ Rev.\ D {\bf 98} (2018) no.3,  030001.

\bibitem{Batley:2002gn}
  J.~R.~Batley {\it et al.} [NA48 Collaboration],
  Phys.\ Lett.\ B {\bf 544} (2002) 97.
  
  
\bibitem{Abouzaid:2010ny}
  E.~Abouzaid {\it et al.} [KTeV Collaboration],
  Phys.\ Rev.\ D {\bf 83} (2011) 092001.
  
\bibitem{Cirigliano:2003gt}
  V.~Cirigliano, G.~Ecker, H.~Neufeld and A.~Pich,
  Eur.\ Phys.\ J.\ C {\bf 33} (2004) 369.
 
  
\bibitem{Antonelli:2010yf}
  M.Antonelli {\it et al.} [FlaviaNet Working Group on Kaon Decays],
  Eur.Phys.J. C {\bf 69} (2010) 399.
  
\bibitem{Gisbert:2018tuf}
  H.~Gisbert and A.~Pich,
  Nucl.\ Part.\ Phys.\ Proc.\  {\bf 300-302} (2018) 137.
 
\bibitem{Colangelo:2000jc}
  G.~Colangelo, J.~Gasser and H.~Leutwyler,
  Phys.\ Lett.\ B {\bf 488} (2000) 261.

  
  
  
\bibitem{Blum:2015ywa}
  T.~Blum {\it et al.},
  Phys.\ Rev.\ D {\bf 91} (2015) no.7,  074502.
  
\bibitem{Bai:2015nea}
  Z.~Bai {\it et al.} [RBC and UKQCD Collaborations],
  Phys.\ Rev.\ Lett.\  {\bf 115} (2015) no.21,  212001.
  
  
\bibitem{Fu:2013ffa}
  Z.~Fu,
  Phys.\ Rev.\ D {\bf 87} (2013) no.7,  074501.
  
\bibitem{Kelly:2019yxg}
  C.~Kelly and T.~Wang,
  PoS LATTICE {\bf 2018} (2019) 277.

\bibitem{Buras:2015yba}
  A.~J.~Buras, M.~Gorbahn, S.~Jäger and M.~Jamin,
  JHEP {\bf 1511} (2015) 202.

\bibitem{Buchalla:1995vs}
  G.~Buchalla, A.~J.~Buras and M.~E.~Lautenbacher,
  Rev.\ Mod.\ Phys.\  {\bf 68} (1996) 1125.

  
\bibitem{Buras:1991jm}
  A.~J.~Buras {\it et al.} 
  Nucl.\ Phys.\ B {\bf 370} (1992) 69
   [Addendum: Nucl.\ Phys.\ B {\bf 375} (1992) 501].
  
\bibitem{Buras:1992tc} 
  A.~J.~Buras, M.~Jamin, M.~E.~Lautenbacher and P.~H.~Weisz,
  {\it Nucl. Phys.} B {\bf 400} (1993) 37.

\bibitem{Buras:1992zv}
  A.~J.~Buras, M.~Jamin and M.~E.~Lautenbacher,
  Nucl.\ Phys.\ B {\bf 400} (1993) 75.

\bibitem{Ciuchini:1993vr}
  M.~Ciuchini, E.~Franco, G.~Martinelli and L.~Reina,
  {\it Nucl. Phys.} B {\bf 415} (1994) 403.
  
\bibitem{Buras:1999st}
  A.~J.~Buras, P.~Gambino and U.~A.~Haisch,
  Nucl.\ Phys.\ B {\bf 570} (2000) 117.

\bibitem{Gorbahn:2004my}
  M.~Gorbahn and U.~Haisch,
  Nucl.\ Phys.\ B {\bf 713} (2005) 291.

\bibitem{Cerda-Sevilla:2016yzo}
  M.~Cerdà-Sevilla {\it et al.}, 
  J.\ Phys.\ Conf.\ Ser.\  {\bf 800} (2017) no.1,  012008.
  
\bibitem{Cirigliano:2011ny}
  V.~Cirigliano, G.~Ecker, H.~Neufeld, A.~Pich and J.~Portol\'es,
  Rev.\ Mod.\ Phys.\  {\bf 84} (2012) 399.
   
\bibitem{Cirigliano:2003nn}
  V.~Cirigliano, A.~Pich, G.~Ecker and H.~Neufeld,
  Phys.\ Rev.\ Lett.\  {\bf 91} (2003) 162001.

\bibitem{Pallante:1999qf}
  E.~Pallante and A.~Pich,
  Phys.\ Rev.\ Lett.\  {\bf 84} (2000) 2568.

\bibitem{Pallante:2000hk}
  E.~Pallante and A.~Pich,
  Nucl.\ Phys.\ B {\bf 592} (2001) 294.
 
\bibitem{Pallante:2001he}
  E.~Pallante, A.~Pich and I.~Scimemi,
  Nucl.\ Phys.\ B {\bf 617} (2001) 441.
  
\bibitem{Aebischer:2019mtr}
  J.~Aebischer, C.~Bobeth and A.~J.~Buras,
  arXiv:1909.05610 [hep-ph].
  
  
    
\end{thebibliography}
\end{document}